\documentclass[runningheads]{llncs}
\usepackage[T1]{fontenc}
\usepackage{graphicx}
\usepackage{xcolor}
\usepackage{threeparttable}
\usepackage{siunitx}
\usepackage[colorlinks=true, allcolors=blue]{hyperref}
\usepackage{csquotes}
\usepackage{booktabs}
\usepackage{comment}
\usepackage{enumitem}
\usepackage{url}

\usepackage{amsmath}
\begin{document}
\pdfinfo{
/Title (Sandwiched and Silent: Behavioral Adaptation and Private Channel Exploitation in Ethereum MEV)
}

\title{Sandwiched and Silent:\\ Behavioral Adaptation and Private Channel Exploitation in Ethereum MEV}
\titlerunning{Sandwiched and Silent}

\author{}
\institute{}

\author{Davide Mancino\inst{1} \and
Davide Rezzoli\inst{2}}

\authorrunning{D. Mancino and D. Rezzoli}

\institute{\emph{University of Milano-Bicocca,
\email{d.mancino1@campus.unimib.it}
  \and
PBS Foundation,
\email{davide@pbs.foundation}}\\
} 

\maketitle              % typeset the header of the contribution
\begin{abstract}
How users adapt after being sandwiched remains unclear; this paper provides an empirical quantification. Using transaction level data from November 2024 to February 2025, enriched with mempool visibility and ZeroMEV labels, we track user outcomes after their $n$-th public sandwich: (i) reactivation, i.e., the resumption of on-chain activity within a 60-day window, and (ii) first-time adoption of private routing. We refer to users who do not reactivate within this window as \emph{churned}, and to users experiencing multiple attacks ($n>1$) as undergoing \emph{repeated exposure}. 
Our analysis reveals measurable behavioral adaptation: around 40\% of victims migrate to private routing within 60 days, rising to 54\% with repeated exposures. Churn peaks at 7.5\% after the first sandwich but declines to 1--2\%, consistent with survivor bias. 
In Nov--Dec 2024 we confirm 2,932 private sandwich attacks affecting 3,126 private victim transactions, producing \$409,236 in losses and \$293,786 in attacker profits. A single bot accounts for nearly two-thirds of private frontruns, and private sandwich activity is heavily concentrated on a small set of DEX pools. 
These results highlight that private routing does not \emph{guarantee} protection from MEV extraction: while execution failures push users toward private channels, these remain exploitable and highly concentrated, demanding continuous monitoring and protocol-level defenses.

\end{abstract}

\section{Introduction}

Maximal Extractable Value (MEV)~\cite{ethereum:mev:onlinedocs:2023} has reshaped incentives and execution risk in Ethereum~\cite{buterin:ethereum:whitepaper:2014}. In Proof-of-Stake Ethereum~\cite{buterin:combiningghostcasper:arxiv:2020}, validators acting as \emph{proposers} ultimately decide block contents and ordering, which exposes the system to MEV, the profit obtainable by manipulating transaction order. To mitigate the systemic risks of specialized infrastructure, power asymmetries, and centralization pressures, Ethereum adopted the \emph{Proposer–Builder Separation} (PBS) paradigm: \textbf{builders} construct candidate blocks and bid for inclusion, while \textbf{proposers} select among these bids, typically choosing the most profitable block. Off-chain, Flashbots’ \emph{MEV-Boost}~\cite{flashbots:github_mevboost:onlinedocs:2022,flashbots:intro_mevboost:onlinedocs:2022} operationalizes this market via trusted relays, reducing infra requirements for proposers but raising concerns around relay centralization and censorship. Ongoing protocol research on \emph{enshrined PBS (ePBS)}~\cite{EIP:ePBS:EIP:2024} aims to internalize the separation, strengthening decentralization and censorship-resistance guarantees.

Within this PBS landscape, \emph{private order flow} emerged as a safety valve. Rather than broadcasting to the public mempool, users route transactions through closed channels, directly to builders, relays, or specialized services to limit pre-trade information leakage and reduce exposure to frontrunning and related MEV strategies. However, private routing is an alternative \emph{path} for order flow, not an absolute shield: inclusion policies, cross-domain leakage, and strategic interactions among intermediaries can still enable adversarial ordering. Notably, \emph{sandwich attacks}, classically associated with public mempool dynamics, also manifest along private paths and at scales that challenge the perceived safety of private routing.

This paper investigates how users adapt after being subjected to sandwich attacks and quantifies the extent to which private routing provides protection. We combine transaction-level data from November~2024 to February~2025 with historical mempool visibility labels 
(via \emph{MempoolDumpster}~\cite{flashbots:mempool_dumpster:onlinedocs:2024}) 
and MEV classifications (via \emph{ZeroMEV}~\cite{zeromev:zeromev:onlinedocs:2024}) 
to study post-attack behavior (reactivation, churn, and migration to private routing) and to characterize the structure of private-path exploitation. Our findings reveal measurable shifts in user behavior alongside critical vulnerabilities within private routing channels, which remain subject to concentrated and systematic exploitation.

\subsection*{Contributions}
This work makes the following contributions:

\begin{enumerate}[label=(\roman*)]
    \item We design a windowed, $n$-indexed survival analysis to measure user reactivation and private routing adoption following the $n$-th public sandwich attack.  
    \item We present the first systematic detection and economic quantification of private-path sandwich attacks, identifying thousands of victim transactions and attacker profits.  
    \item We show that migration to private routing does not prevent MEV extraction: 3,126 private victim transactions (Nov–Dec 2024) demonstrate systematic exploitation.  
    \item We document extreme concentration among attackers: one operator controls approximately 65\% of private sandwich volume, suggesting potential compromise or collusion within private routing infrastructure.  
\end{enumerate}

Taken together, these findings challenge the assumption that private routing offers genuine protection from MEV. 
Instead, they reveal that private channels may constitute a narrower but highly concentrated attack surface, demanding ongoing monitoring and protocol-level defenses.

The remainder of the paper is organized as follows.
Section~\ref{sec:related_work} reviews prior research on MEV, PBS, and private order flow.
Section~\ref{sec:data} describes our dataset, visibility labeling, and MEV annotation.
Section~\ref{sec:method} details the sandwich-detection methodology.
Section~\ref{sec:findings} reports empirical results on user adaptation and private-path exploitation and section~\ref{sec:conclusion} concludes.

\section{Related Work}\label{sec:related_work}
Foundational studies showed how discretionary ordering enables extractable value and harms market fairness: Daian et al.\ formalized MEV on DEXs and its consensus externalities \cite{daian:flashboys20:sp:2020}; Piet et al.\ analyzed miner-side extraction from DeFi order flow \cite{piet:extracting_godl:arxiv:2022}; Qin et al.\ quantified system-wide impacts and called for principled mitigations \cite{qin:quantifying_bev:sp:2022}. With Ethereum’s move to Proof-of-Stake, the locus of ordering power shifted. Measurements around PBS document concentration among builders/relays, mixed relay reliability, and patterns consistent with censorship under PBS compared to non-PBS blocks \cite{heimbach:promises_pbs:imc:2023}; timing frictions remain central to rewards \cite{oz:time_moves_faster:defi:2023}; cross-consensus comparisons clarify how protocol choices shape MEV and fairness \cite{oz:algorand:icbc:2024}. An Ethereum PoS study over January 2024 (220{,}993 blocks; 36{,}015{,}340 txs) reports systematic prioritization of private flows (especially when \emph{builder payments} are present) and relative delays for public mempool transactions, linking these effects to speculative strategies and MEV-bot interactions \cite{mancino:decentralization_favoritism:icbc:2025};

Private order-flow infrastructure predates PBS: empirical work on Flashbots during PoW observes near-complete adoption among major miners, large revenue uplifts, and high centralization, with most MEV routed via private relays \cite{weintraub:flashbot:imc:2022}. Beyond L1, rollups exhibit widespread extractive practices with different cost/profit profiles; classical sandwiching is muted by centralized sequencing and the absence of public mempools, yet cross-layer sandwich vectors bridging L1–L2 are feasible and historically profitable in counterfactuals \cite{torres:rolling_shadows_mev:ccs:2024}. A measurement on FCFS rollups finds that profit-maximizing bots split opportunities post-Dencun; trace-level graphs across major L2s show that >80\% of reverts are swaps ($\approx$half USDC–WETH on Uniswap v3/v4), reverts cluster near block tops, and duplicate submissions outweigh priority-fee bidding, underscoring the fragility of fee-based ordering at sub-second block times \cite{gogol:firstspammedfirstservedmevextraction:arxiv:2025}.
Theory indicates heterogeneous rewards in PBS-style markets can still amplify builder power and stake concentration \cite{bahrani:centralization_block_building:springer_fc:2025}; at the DEX layer, game-theoretic models motivate adaptive slippage that significantly reduces sandwich losses while private bundles lower attacker execution risk \cite{heimbach:game_theory_sandwich:asiaccs:2022}.

Our contribution situates here by examining how sandwiched users adapt over time and by documenting that \emph{private} transactions, not only public mempool ones, are themselves subject to sandwich attacks.

\section{Data Collection}
\label{sec:data}

We construct a transaction-level panel spanning November~2024 to February~2025, distinguishing public and private visibility and enriching each hash with high-level MEV semantics. 

Raw transaction metadata are sourced from \textbf{XBlock-ETH}~\cite{zheng:xblock_eth:ieee:2020} at block granularity (fields: \texttt{blockNumber}, \texttt{timestamp},   \texttt{transactionHash}, \texttt{from}, \texttt{to}). The analysis covers the following block ranges:  November 2024 \allowbreak(21089069 -- 21303933), December 2024 (21303934 -- 21525890), January 2025 (21525891 -- 21747949), and February 2025 (21747950 -- 21948291).

\paragraph{Visibility and MEV labels.}
A transaction is labeled \textbf{public} if observed in the mempool prior to inclusion, \textbf{private} otherwise, using \textbf{MempoolDumpster}~\cite{flashbots:mempool_dumpster:onlinedocs:2024} for historical mempool cross-referencing. For each hash, the \textbf{ZeroMEV} API~\cite{zeromev:zeromev:onlinedocs:2024} provides \texttt{mev\_type} $\in$ \{\texttt{swap}, \texttt{none}, \texttt{arbitrage}, \texttt{liquidation}, \texttt{frontrun}, \texttt{backrun}, \texttt{sandwich} (victim of the attack)\}, the target \texttt{protocol} (e.g., \texttt{uniswap2}, \texttt{uniswap3}), monetized summaries (user loss, extractor profit), and swap volumes/counts.

\begin{figure}[ht!]
    \centering
    \includegraphics[width=0.9\textwidth]{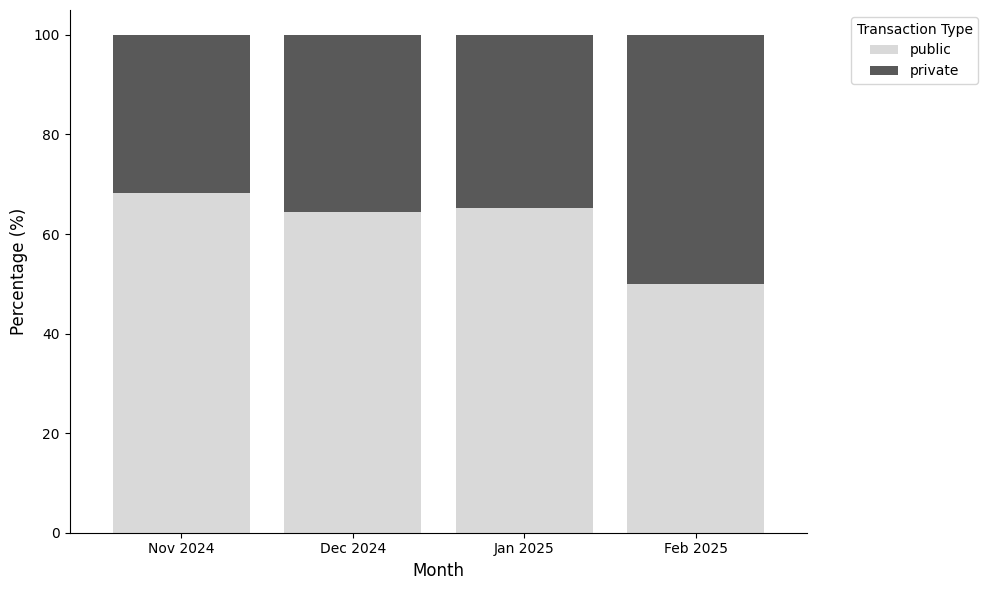}
    \caption{Monthly distribution of public vs.\ private transactions (Nov 2024 -- Feb 2025).}
    \label{fig:pub_priv}
\end{figure}

\paragraph{Monthly overview, visibility split, and MEV breakdown.}
Tables~\ref{tab:overview_4m} and~\ref{tab:visibility_4m} consolidates totals and the public/private split across Nov 2024--Feb 2025. Figure~\ref{fig:pub_priv} shows the month-over-month rise in private routing: private share is $\sim$31.8\% in November and $\sim$35.5\% in December, \textbf{34.8\%} in January (13{,}189{,}288 / 37{,}865{,}791), and reaches \textbf{50.1\%} in February (17{,}807{,}840 / 35{,}574{,}446), marking a shift from public-dominated to near-balanced visibility. For the MEV analysis, detailed ZeroMEV labels were collected for \textbf{November and December 2024}; accordingly, Table~\ref{tab:visibility_x_mev_nov_dec} reports the combined \emph{visibility $\times$ MEV} counts for those two months.

\begin{table}[ht!]
\centering
\small
\caption{Overview of Ethereum transactions (Nov 2024 -- Feb 2025).}
\label{tab:overview_4m}
\begin{tabular}{p{3cm}rrrr}
\toprule
 & \multicolumn{1}{c}{Nov 2024} & \multicolumn{1}{c}{Dec 2024} & \multicolumn{1}{c}{Jan 2025} & \multicolumn{1}{c}{Feb 2025} \\
\midrule
Total transactions & 36,691,046 & 39,605,073 & 37,865,791 & 35,574,446 \\
Blocks             & 214,865    & 221,957    & 222,059    & 200,342 \\
\bottomrule
\end{tabular}
\end{table}

\begin{table}[ht!]
\centering
\small
\caption{Visibility classification by month (Nov 2024 -- Feb 2025).}
\label{tab:visibility_4m}
\begin{tabular}{p{3cm}rrrr}
\toprule
 & \multicolumn{1}{c}{Nov 2024} & \multicolumn{1}{c}{Dec 2024} & \multicolumn{1}{c}{Jan 2025} & \multicolumn{1}{c}{Feb 2025} \\
\midrule
Public  & 25,010,079 & 25,551,930 & 24,676,503 & 17,766,606 \\
Private & 11,680,967 & 14,053,143 & 13,189,288 & 17,807,840 \\
\bottomrule
\end{tabular}
\end{table}

\begin{table}[ht!]
\centering
\footnotesize
\caption{Combined visibility $\times$ MEV classification (counts) by month (Nov--Dec 2024).}
\begin{threeparttable}
\label{tab:visibility_x_mev_nov_dec}
\begin{tabular}{p{2.5cm}p{2.5cm}rrr}
\toprule
\multicolumn{1}{l}{Visibility} & \multicolumn{1}{l}{MEV Type} & & \multicolumn{1}{c}{Nov 2024} & \multicolumn{1}{c}{Dec 2024} \\
\midrule
Public  & None        & & 22,990,098 & 23,914,461 \\
Private & None        & & 6,830,414  & 9,774,836 \\
Private & Swap        & & 4,618,825  & 4,134,442 \\ 
Public  & Swap        & & 1,931,760  & 1,584,404 \\
Private & Backrun     & & 88,956     & 53,974 \\
Private & Frontrun    & & 88,842     & 53,965 \\
Public  & Sandwich    & & 86,906     & 50,980 \\
Private & Arbitrage   & & 43,075     & 26,248 \\
Private & Sandwich    & & 10,609     & 9,204 \\
Public  & Backrun     & & 538        & 944 \\
Public  & Frontrun    & & 529        & 912 \\
Private & Liquidation & & 246        & 474 \\
Public  & Arbitrage   & & 245        & 217 \\
Public  & Liquidation & & 3          & 12 \\
\bottomrule
\end{tabular}
\end{threeparttable}
\end{table}

\section{Methodology: Sandwich Attack Detection}
\label{sec:method}

We detect sandwiches on a ZeroMEV-enriched dataset, enforcing \emph{block-consistent} ordering and confirming only legitimate triplets.

To detect sandwich attacks, we begin by sorting all transactions by \textit{block number} and \textit{transaction index}, ensuring a consistent execution order within each block. We then initialize a \textit{sandwich-role} field by copying existing MEV labels provided by ZeroMEV.

Within each block, we scan for frontrun candidates. For each transaction at index $i$ labeled as a frontrun and sent by address $s$, we locate the first subsequent transaction (index $j > i$) sent by the same address $s$ and labeled as a backrun. We consider the sequence from $i+1$ to $j-1$ as the candidate victim region.

From this region, we extract transactions that ZeroMEV labels as \textit{sandwich} (in ZeroMEV, this label denotes the \emph{victim} of a sandwich attack) identifying them as potential victims. A sandwich is confirmed if at least one valid victim exists between the frontrun and backrun. When confirmed, the corresponding frontrun–backrun pair is marked accordingly.

To ensure integrity, we apply a set of consistency checks: we exclude self-sandwiching by requiring that the victim's sender is different from the attacker ($s$), and we retain only distinct external victims.

Finally, if any confirmed victim transaction is labeled as \textit{private} (i.e., not observed in the public mempool), we record the associated frontrun sender as a private-path attacker. This allows us to systematically track the exploitation of private order flow across blocks.

The final set includes only validated sandwiches where (i) frontrun and backrun share the same sender, (ii) at least one external victim lies between them within the block, and (iii) private victims are linked to their attackers, enabling analyses by visibility and attacker capability.

\section{Empirical Findings}
\label{sec:findings}

After applying all filtering and integrity checks, we obtain the following totals:
\emph{public} victims: \textbf{123{,}030} events; \emph{private} victims: \textbf{3{,}126} events.
We confirm \emph{public} sandwich attacks (paired frontrun{+}backrun): \textbf{117{,}581}, and \emph{private} sandwich attacks (paired frontrun{+}backrun): \textbf{2{,}932}.

\subsection{Distribution of Sandwich Events}
\begin{figure}[ht]
\centering
\includegraphics[width=0.7\linewidth]{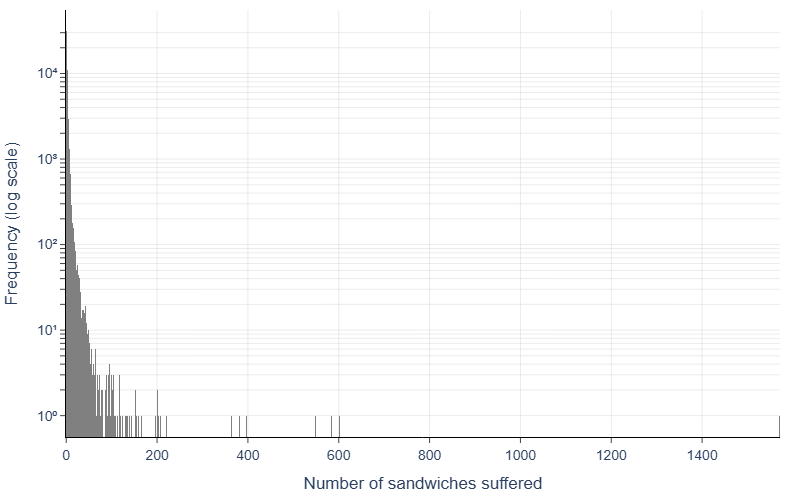}
\hfill
\includegraphics[width=0.7\linewidth]{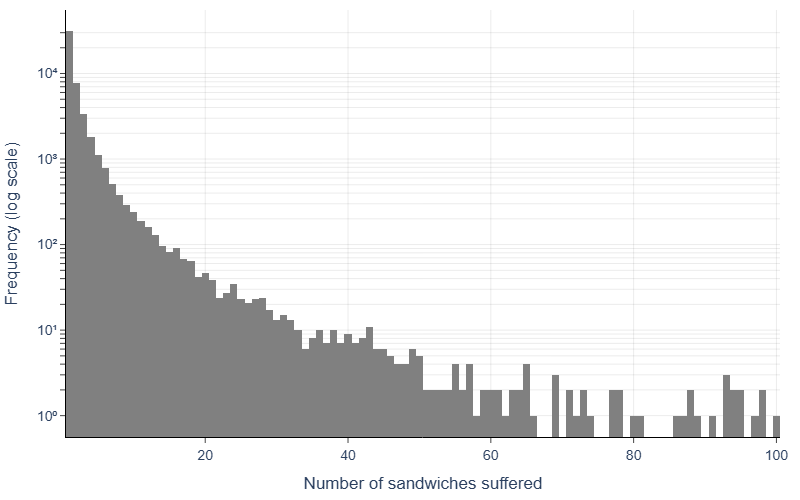}
\caption{Distribution of sandwiches per address. Top: full range (long tail extends beyond 100). Bottom: zoom for $n \in [1,100]$, highlighting the dense head and gradual decline.}
\label{fig:sandwich_distribution}
\end{figure}

Restricting to addresses that are ever sandwiched, the majority suffer only 1--3 events. The top panel of Figure~\ref{fig:sandwich_distribution} shows a heavy-tailed distribution extending well beyond 100 events, while the bottom panel (zoom for $n\in[1,100]$) highlights the dense head, where frequencies decline gradually with $n$.

\begin{figure}[ht]
\centering
\includegraphics[width=0.9\linewidth]{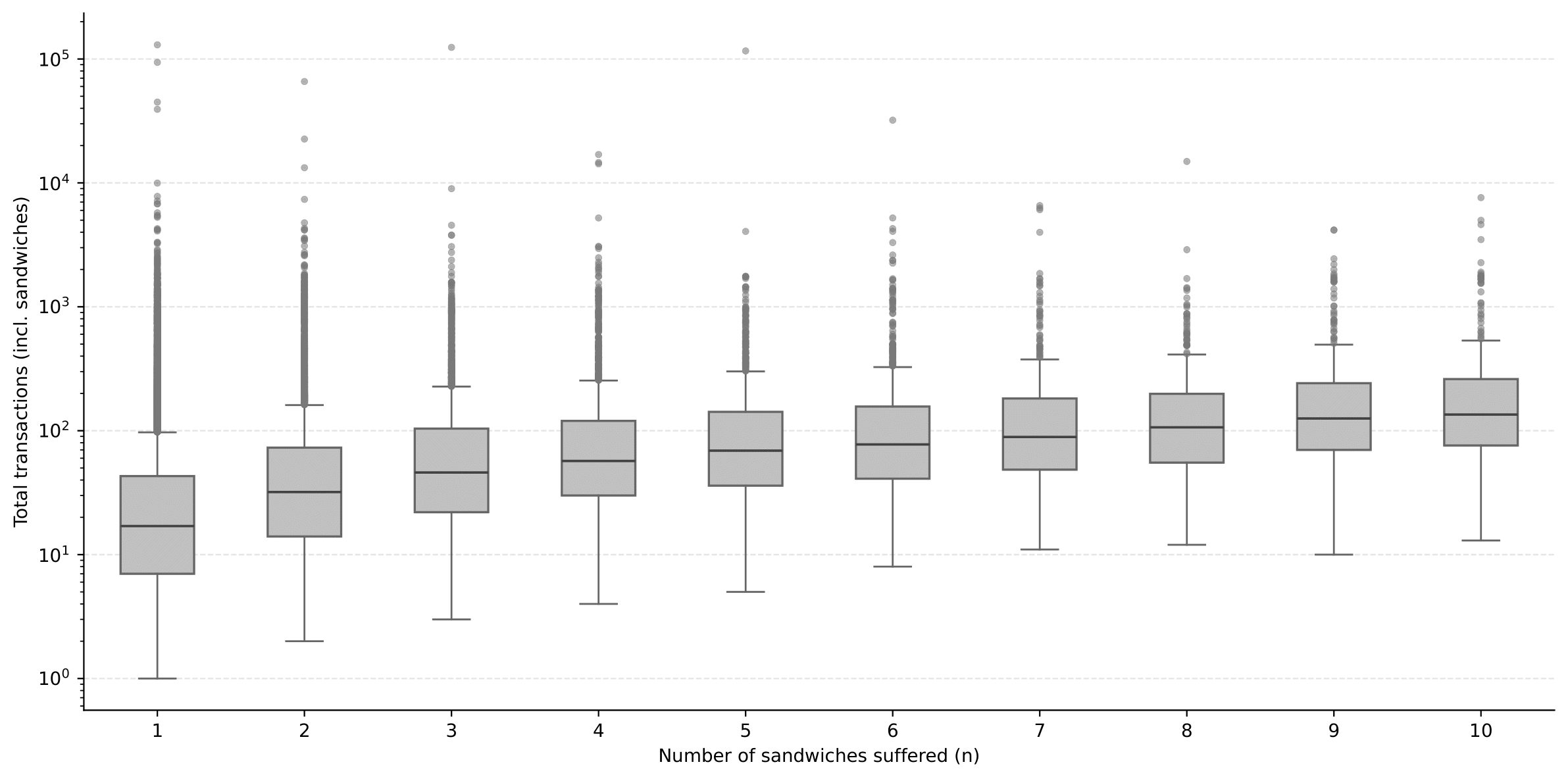}
\caption{Total transactions (including sandwiches) by number of sandwiches suffered.}
\label{fig:dist_txs_x_n_sandwich}
\end{figure}

Moreover, as the number of sandwiches received increases, overall activity also rises: Figure~\ref{fig:dist_txs_x_n_sandwich} shows that addresses with higher sandwich exposure tend to account for more total transactions.

\subsection{User Reactivation Analysis and Survival Patterns}

We investigate the long-term behavioral impact of sandwich attacks on user activity through an innovative adaptation of survival analysis methodologies. Our approach inverts traditional Kaplan-Meier~\cite{kaplanmeier:kaplanmeier:ASA:1958} survival concepts by examining user reactivation probabilities rather than activity cessation, providing novel insights into how MEV exploitation affects on-chain participation.

Our methodology tracks addresses that have experienced a maximum of $n$ sandwich attacks, monitoring their subsequent activity patterns within a 60-day observation window following their $n$-th attack exposure. We define reactivation as any transaction executed by the targeted address after experiencing the $n$-th sandwich attack, creating a binary outcome measure that captures continued on-chain engagement despite MEV exploitation.

\begin{figure}[ht]
\centering
\includegraphics[width=0.99\linewidth]{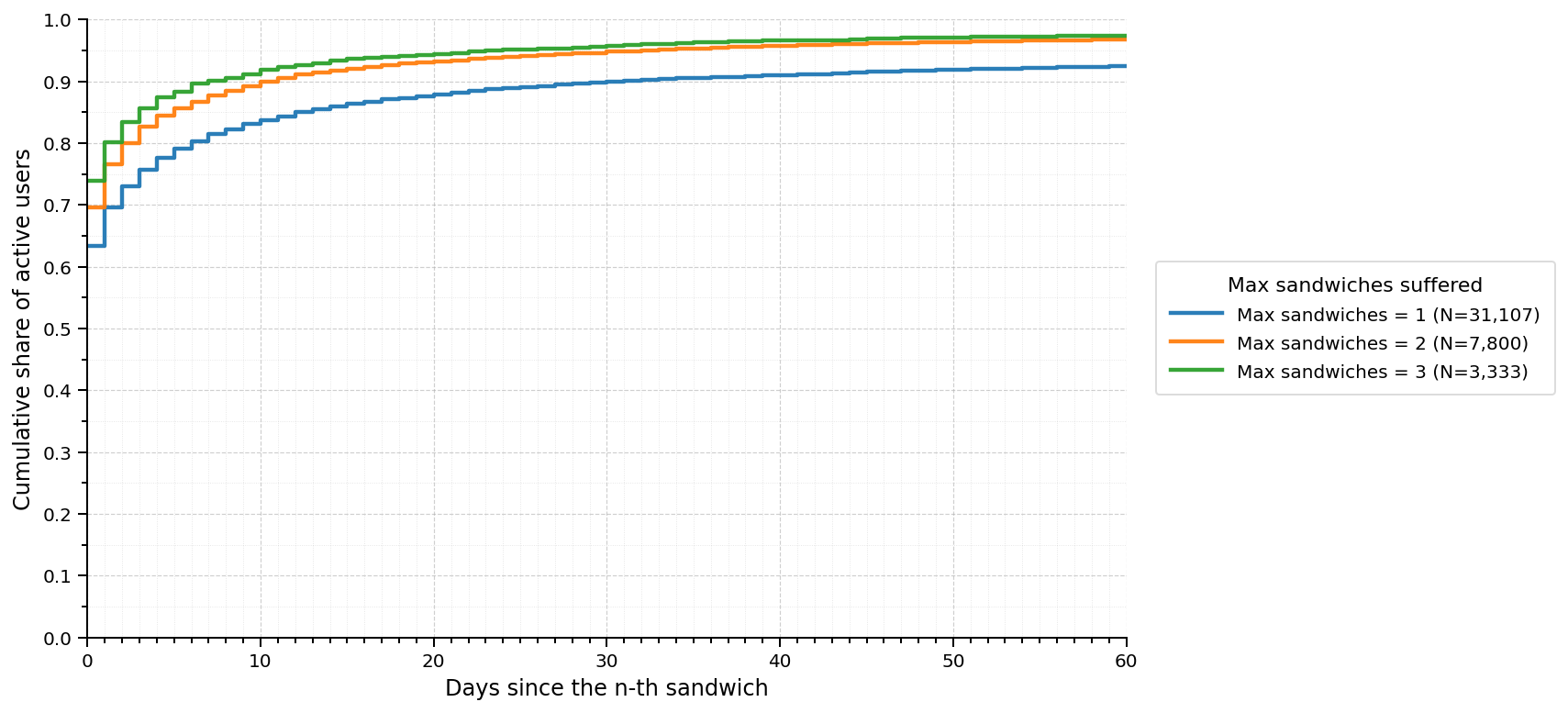}
\hfill
\includegraphics[width=0.99\linewidth]{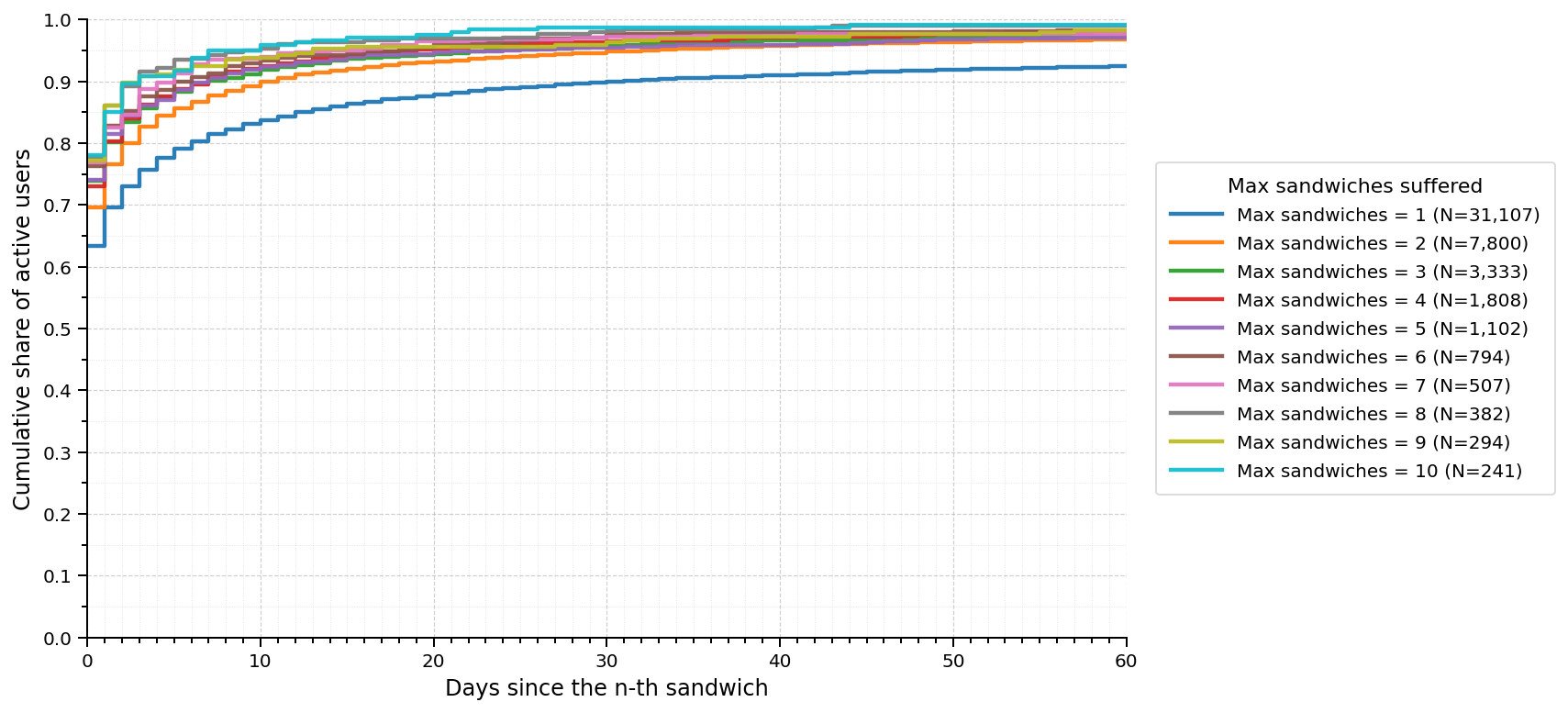}
\caption{Cumulative incidence of reactivation and private adoption. Top: addresses subjected to up to 3 sandwich attacks; bottom: addresses subjected to up to 10.}
\label{fig:sopravvivenza}
\end{figure}

The temporal dynamics of reactivation behavior, illustrated in the cumulative incidence curves of Figures~\ref{fig:sopravvivenza}, reveal surprising patterns that challenge conventional assumptions about user response to adversarial activity. A substantial proportion of addresses resume activity on the same day as their sandwich attack, with many reactivating within hours of exploitation. This immediate reactivation behavior strongly suggests that the majority of users remain completely oblivious to MEV extraction or have incorporated such costs into their expected transaction overhead.

The cumulative curves demonstrate rapid reactivation trajectories, with the steepest increases in the share of reactivated addresses occurring within the first 24–48 hours post-attack. Users subjected to multiple attacks exhibit even more pronounced immediate recovery patterns, with nearly instantaneous resumption of trading activity becoming increasingly common as attack frequency rises. This behavioral adaptation indicates either sophisticated users who accept MEV as an inevitable transaction cost or naive users who remain systematically unaware of the exploitation.

\begin{figure}[ht!]
\centering
\includegraphics[width=0.8\linewidth]{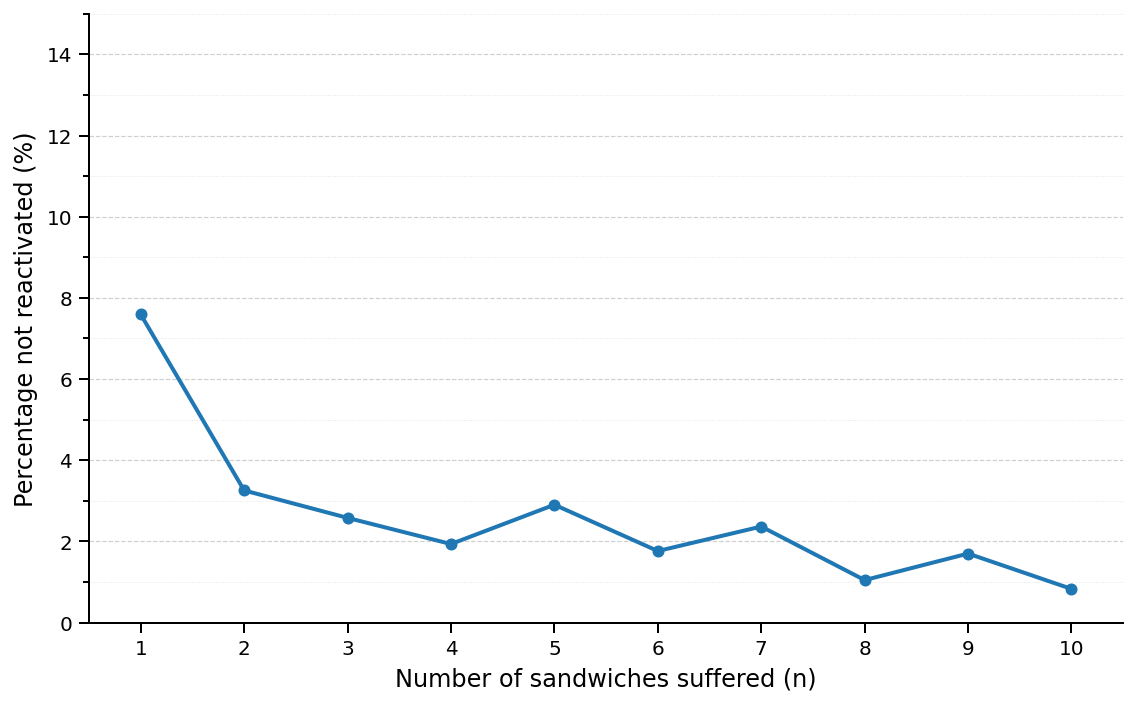}
\caption{Share of users \emph{not} reactivated within 60 days by sandwich index $n$. Churn risk is highest at $n=1$ and decreases with repeated exposure.}
\label{fig:non_sopravvissuti}
\end{figure}

These temporal patterns become even more striking when examined alongside the aggregate reactivation statistics. Figure~\ref{fig:non_sopravvissuti} demonstrates that churn risk exhibits a pronounced front-loaded distribution, with users experiencing their first sandwich attack showing the highest non-reactivation rates at approximately 7.5\%. This elevated abandonment rate following initial exposure suggests that some users may detect the exploitation and choose to cease their trading activities or migrate to alternative platforms.

However, as sandwich attack frequency increases, non-reactivation rates decrease dramatically, stabilizing around 1-2\% for users who have endured multiple attacks. This declining trend indicates that addresses surviving the initial MEV exposure remain unaware of the ongoing value extraction or, if aware, choose to continue operating despite it. 

The data reveal a clear survivor bias effect among frequently targeted addresses: users who have endured up to ten sandwich attacks show reactivation rates exceeding 98\%. This population represents addresses that demonstrate exceptional tolerance toward MEV extraction, potentially reflecting systematic differences in user sophistication or awareness levels within the DeFi ecosystem. 

The paradoxical relationship between attack frequency and user retention suggests that successful MEV extraction creates a self-reinforcing cycle, where addresses that continue to operate after their first attacks become increasingly attractive targets for repeated exploitation. The highest attrition occurs among users subjected to isolated attacks, potentially representing more vigilant participants who recognize the predatory behavior and subsequently adjust their engagement patterns.

\subsection{Private Routing Adoption Patterns}
\label{sec:private_routing}

The adoption of private transaction routing represents a sophisticated behavioral adaptation to sandwich attack exposure, revealing how users evolve their trading strategies in response to MEV exploitation. Our analysis examines the propensity of users to migrate from public mempool submission to private routing mechanisms following sandwich attacks, providing insights into defensive behavior patterns across the DeFi ecosystem.

\begin{figure}[ht!]
\centering
\includegraphics[width=0.48\linewidth]{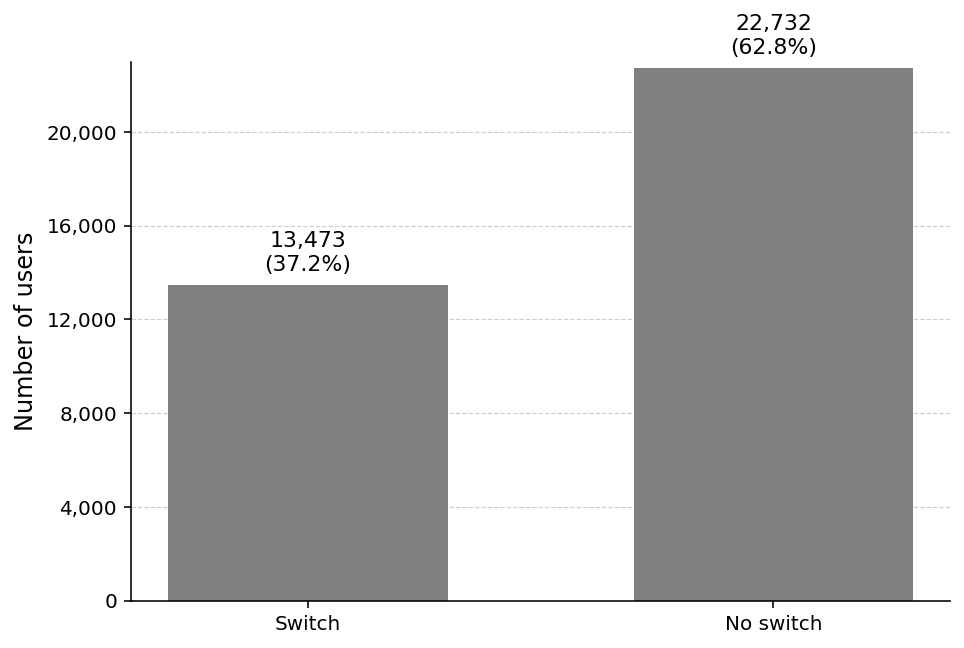}
\hfill
\includegraphics[width=0.48\linewidth]{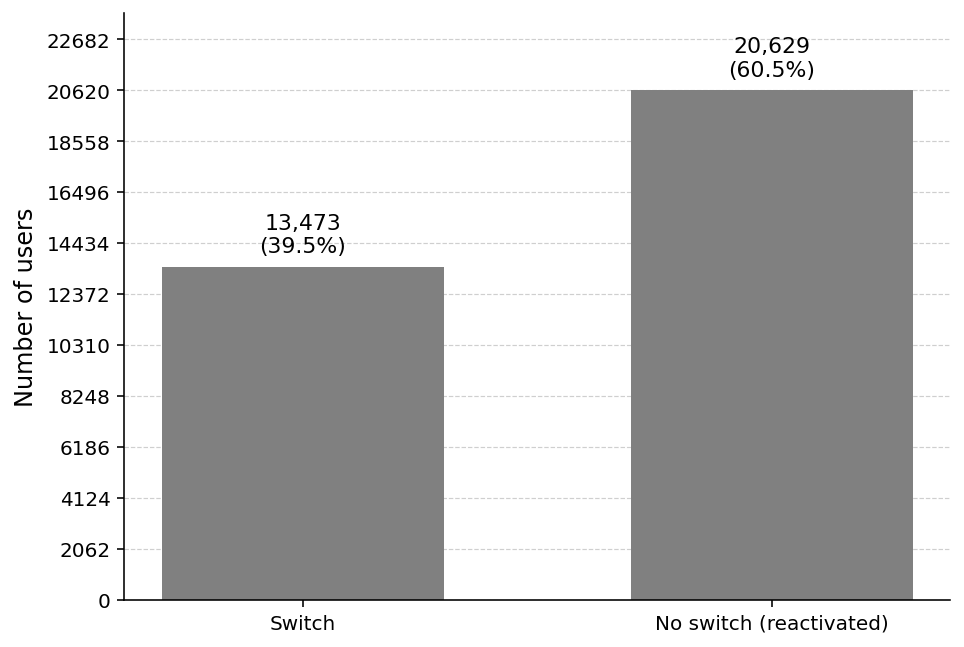}
\caption{Private routing adoption following the first public sandwich attack. Left panel shows all users in the dataset (N=36,205), while the right panel excludes users who did not reactivate within 60 days, focusing on active participants (N=34,102). The conditional analysis reveals higher adoption rates among users who continue trading activity.}
\label{fig:switch}
\end{figure}

Figure~\ref{fig:switch} demonstrates that private routing adoption following the first public sandwich attack is far from negligible, with 13,473 users (37.2\%) transitioning to private channels within 60 days of their initial attack exposure. This substantial migration rate indicates that a significant portion of the user base possesses both the awareness and technical sophistication necessary to implement protective measures against MEV extraction.

The behavioral patterns become even more pronounced when we examine users who demonstrate continued engagement with the platform. Among addresses that reactivate within the 60-day observation window, private routing adoption increases to 39.5\% (13,473 out of 34,102 reactivated users), as shown in the conditional analysis. This elevated adoption rate among active users suggests that continued platform engagement correlates with increased awareness of MEV risks and available mitigation strategies.

\begin{figure}[ht!]
\centering
\includegraphics[width=0.65\linewidth]{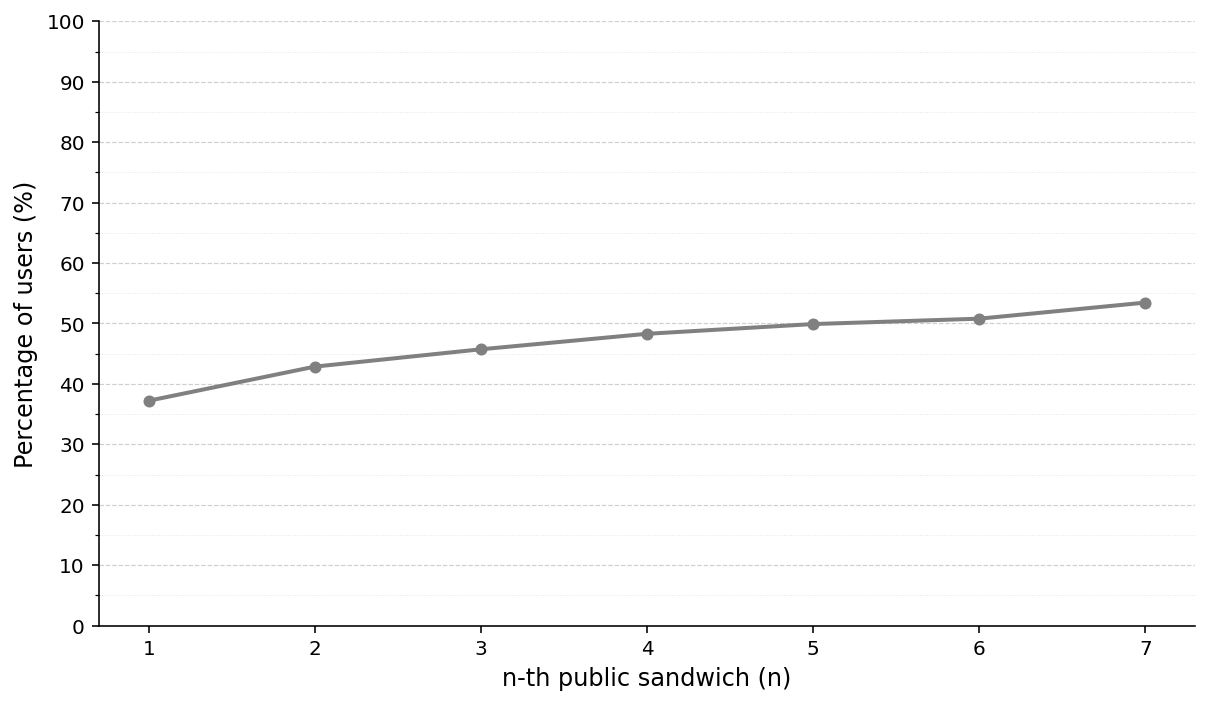}
\hfill
\includegraphics[width=0.65\linewidth]{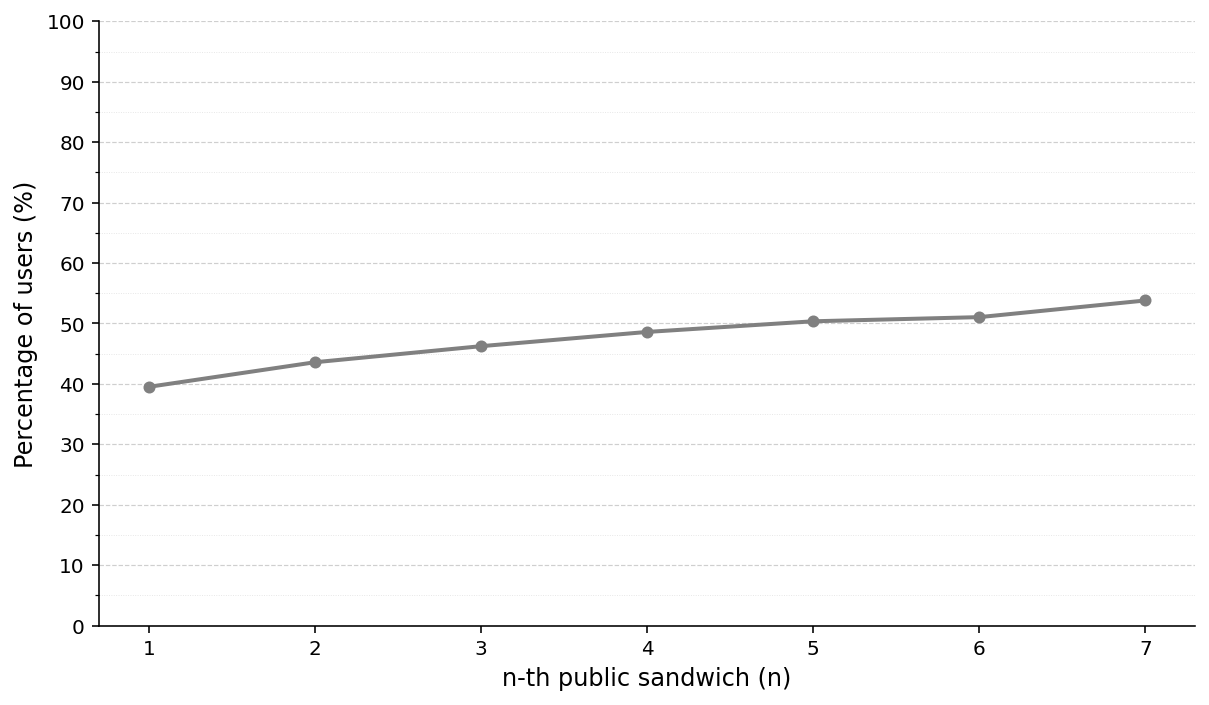}
\caption{Private routing adoption rates across multiple sandwich exposures ($n \in [1,7]$). The top panel includes all users experiencing their $n$-th public sandwich attack, while the bottom panel excludes non-reactivated addresses from the analysis population. Both panels demonstrate steadily increasing adoption rates with repeated exposure, indicating cumulative learning effects.}
\label{fig:switch_1_7}
\end{figure}

The progression of private routing adoption across multiple sandwich exposures reveals a clear learning effect that strengthens with repeated victimization. Figures~\ref{fig:switch_1_7} illustrate a steady upward trend in adoption rates as users experience additional sandwich attacks. Among all users, the adoption rate increases gradually from approximately 37\% after the first attack to over 54\% by the seventh attack, demonstrating cumulative learning effects that drive protective behavior adoption.

This learning pattern becomes even more pronounced when focusing exclusively on users who maintain active trading behavior. Among reactivated users, private routing adoption follows a similar upward trajectory, reaching comparable peak levels while maintaining consistently higher rates throughout the exposure sequence. The parallel trends across both populations confirm that the learning effect operates independently of user retention patterns.

\begin{figure}[ht!]
\centering
\includegraphics[width=0.9\linewidth]{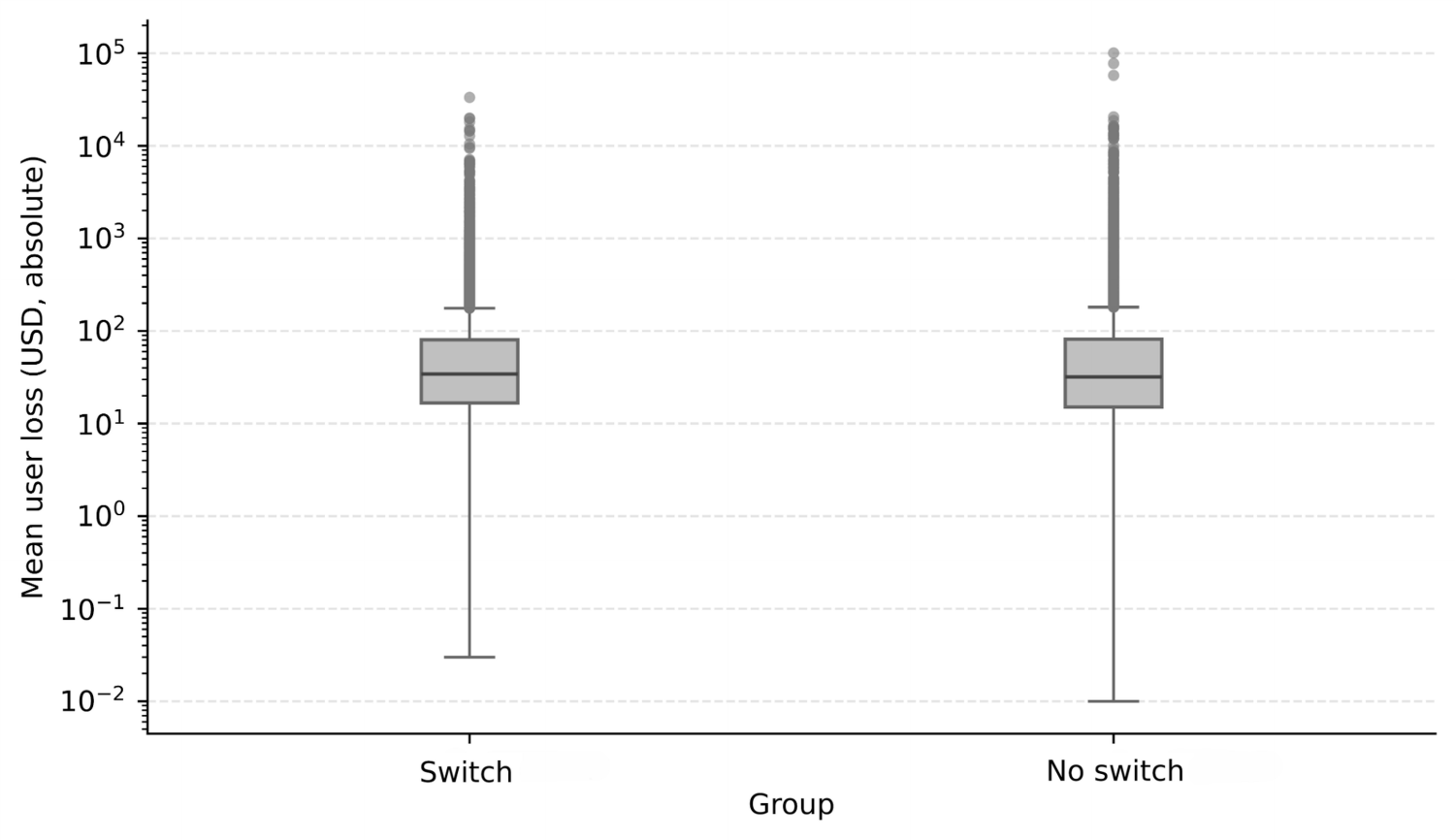}
\caption{Distribution of absolute user losses (USD) comparing addresses that switch to private routing versus those that maintain public mempool usage.}
\label{fig:switch_loss_box}
\end{figure}

The economic context surrounding private routing adoption reveals nuanced patterns when examining loss distributions across different user populations. The boxplot analysis in Figure~\ref{fig:switch_loss_box} demonstrates that switchers and non-switchers experience broadly similar loss distributions. Consistent with this, a \emph{Mann-Whitney U test} \cite{mannwhitney:testmann-whitney:annalsMathematicalStatistics:1947,fay:testmann-whitney:statisticssurveys:2010} detects a difference due to the large sample, but the effect is negligible: Cliff’s $\delta \approx$ 0.036, implying practically indistinguishable loss profiles.

Hence, the decision to adopt private routing does not appear to be explained by systematically higher economic damage. Instead, adoption may be influenced by heterogeneous, unobserved factors at the individual level, such as perceived risk or protective behavior.

The gradual increase in adoption rates across successive sandwich exposures indicates that private routing migration represents a learning process rather than an immediate reaction to attack detection. This delayed response pattern suggests that users require multiple exposures to develop both awareness of MEV risks and the technical knowledge necessary to implement private routing solutions. The cumulative learning effect demonstrates that switching behavior emerges through repeated interactions with the MEV ecosystem, where users progressively recognize the benefits of protective mechanisms regardless of their individual loss experiences.

\subsection{Private-Path Sandwiching}
In Nov--Dec 2024, we confirm \textbf{2,932} private sandwich attacks over \textbf{3,126} private victim transactions (counts differ because a single attack can sandwich multiple victims within the same attack), corresponding to victim-side losses of \(\$409{,}236.97\) and attacker-side profits of \$293{,}785.95. Losses are heavy-tailed, with mean \(\$137.47\), median \(\$26.85\), interquartile range \([\$12.09, \$69.16]\), and a maximum of \$2,744.69. Profits show a similarly skewed distribution, with mean \$106.95, median \$16.35, interquartile range \([\$7.47,\,\$43.05]\), and a maximum of \$37,965.21. Overall, private-path sandwiching exhibits the typical MEV pattern of modest median values combined with occasional extreme extractions.

Concentration patterns among attackers reveal extreme centralization within the private sandwich ecosystem. As demonstrated in Table~\ref{tab:top_private_attackers}, a single bot (address 0xae2f...ae13) executes 1,901 private frontruns, accounting for nearly 65\% of the total attack volume. Etherscan records activity involving the ENS name \texttt{jaredfromsubway.eth} for this address; we report this for context while noting that block-explorer tags and ENS associations may evolve over time. 

The remaining frontruns distribute across a long tail of smaller operators, with the second-largest attacker responsible for only 229 events. This concentration suggests that private-path sandwich attacks require sophisticated technical infrastructure and market access that remains accessible to only a limited number of operators.

\begin{table}[ht]
\centering
\small
\caption{Top attackers by private frontrun count (Nov--Dec 2024).}
\label{tab:top_private_attackers}
\begin{tabular}{lr}
\toprule
Attacker address & Frontrun tx count \\
\midrule
\texttt{0xae2fc483527b8ef99eb5d9b44875f005ba1fae13} & 1,901 \\
\texttt{0x77ad3a15b78101883af36ad4a875e17c86ac65d1} & 229 \\
\texttt{0xa8d7bf2a32d8949e438961fd70505216bb356c01} & 82 \\
\texttt{0xe75ed6f453c602bd696ce27af11565edc9b46b0d} & 67 \\
\texttt{0x00000000000947821264914ad2c75f871aa2d026} & 53 \\
\texttt{0xfc9928f6590d853752824b0b403a6ae36785e535} & 41 \\
\texttt{0x654fae4aa229d104cabead47e56703f58b174be4} & 29 \\
\texttt{0x9b463ee8f0179d60b31b7e1711304d1afb0a6ad6} & 28 \\
\texttt{0xe14386cc2119039d33015be49e4cf0496d740565} & 25 \\
\texttt{0xa85b93983767729fcceac60e32a0bf87cd632108} & 24 \\
\bottomrule
\end{tabular}
\end{table}

The targeting patterns reveal systematic exploitation of certain users, with victim recurrence indicating sustained predatory behavior. Table~\ref{tab:top_private_victims} shows that one address experiences 28 separate private sandwich attacks, followed by others with 23 and 22 victimizations respectively. This recurrence pattern highlights that a subset of users becomes systematically exploited.

\begin{table}[ht]
\centering
\small
\caption{Top victims by private sandwich event count (Nov--Dec 2024).}
\label{tab:top_private_victims}
\begin{tabular}{lr}
\toprule
Victim address & Event count \\
\midrule
\texttt{0xa568bd1f7038bdea2fca05881168eef8fd4ffa33} & 28 \\
\texttt{0xa12e1462d0ced572f396f58b6e2d03894cd7c8a4} & 23 \\
\texttt{0x784ff49d60259924b28236dad62b5dcc309fdb32} & 22 \\
\texttt{0x48b70bfad351ef4af6e6b76d2433535089b0341c} & 16 \\
\texttt{0xef42750eb260c851dc7f30716b3e43ba42299ce8} & 14 \\
\texttt{0x2581aaa94299787a8a588b2fceb161a302939e28} & 13 \\
\texttt{0x19879debeb2c35c630a22d88d89debb12d37c915} & 12 \\
\texttt{0xcfe9830d79ccf00950f5b59fcce65d494c1d856e} & 12 \\
\texttt{0x9e4738c8afda4c1017c7a9410f2f910cc3571a04} & 11 \\
\texttt{0x2b1930524bd624fb6d9387646547f43edd42c826} & 11 \\
\bottomrule
\end{tabular}
\end{table}

Private sandwich attacks overwhelmingly target router and aggregator contracts, with Table~\ref{tab:top_private_destinations} revealing that destination address 0x3fc9...7fad concentrates 856 victim transactions (27\% of total). The aggregator contract 0x1111...0582 accounts for 467 events, while the Uniswap V2 router 0x7a25...488d experiences 397 attacks. The targeting landscape also includes specialized trading interfaces such as Banana Gun (0x3328...9c49), which accounts for 48 victim transactions, demonstrating that private sandwich attacks extend beyond traditional DEX infrastructure to encompass emerging trading tools and user interfaces. Overall, private-path sandwiching appears strongly concentrated on a small set of router and aggregator contracts, rather than being uniformly spread across all possible destinations.

\emph{Robustness to forked blocks.} A natural concern is that some “private” transactions might in fact have become publicly visible if they were first included in blocks that were later orphaned and reported as forked, making their contents observable via public fork listings. To check that our private-path sandwiches are not driven by this effect, we collected all forked blocks over our Nov–Dec 2024
block range from Etherscan’s public forked-block view and intersected them with the set of blocks containing private sandwich victims. Out of all private victim transactions identified in this period, only 8 fall in blocks that are later marked as forked, i.e., well below 0.3\% of the private-sandwich sample. Removing these observations does not change any of the aggregate counts or distributional patterns reported above, suggesting that fork-induced visibility leakage plays a negligible role in our measurements of private-path sandwiching.

\begin{table}[ht]
\centering
\small
\caption{Top destination contracts of private victim transactions (Nov--Dec 2024).}
\label{tab:top_private_destinations}
\begin{tabular}{lr}
\toprule
Destination contract & Victim tx count \\
\midrule
\texttt{0x3fc91a3afd70395cd496c647d5a6cc9d4b2b7fad} & 856 \\
\texttt{0x1111111254eeb25477b68fb85ed929f73a960582} & 467 \\
\texttt{0x7a250d5630b4cf539739df2c5dacb4c659f2488d} & 397 \\
\texttt{0x0000000000001ff3684f28c67538d4d072c22734} & 356 \\
\texttt{0xf3de3c0d654fda23dad170f0f320a92172509127} & 132 \\
\texttt{0xf9e037dcf792ba8c4a0ca570eac7cbcbafabd9d4} & 126 \\
\texttt{0x881d40237659c251811cec9c364ef91dc08d300c} & 100 \\
\texttt{0xb300000b72deaeb607a12d5f54773d1c19c7028d} & 63 \\
\texttt{0x3328f7f4a1d1c57c35df56bbf0c9dcafca309c49} & 48 \\
\texttt{0x7d0ccaa3fac1e5a943c5168b6ced828691b46b36} & 48 \\
\bottomrule
\end{tabular}
\end{table}

\section{Conclusion}
\label{sec:conclusion}

This study provides a large-scale empirical view of MEV extraction in Ethereum’s post-PBS environment, linking user behavior, routing choices, and private-path vulnerabilities. We show that private routing, while increasingly adopted, does not offer reliable protection against sandwich attacks and can concentrate exploitable order flow.

Between November and December 2024 we confirm \textbf{2,932} private sandwich attacks (matched frontrun and backrun pairs) affecting \textbf{3,126} private victim transactions, with aggregate user losses of \textbf{\$409{,}236.97} and attacker profits of \textbf{\$293{,}785.95}. Targeting is highly skewed: private sandwiching is dominated by a narrow set of DEX pools, and a single operator (0xae2f...ae13) accounts for about \textbf{65\%} of private frontruns, pointing to strong centralization in private-path exploitation.

User-level outcomes reveal adaptation, but often in ways that leave users exposed. After a public sandwich, \textbf{37.2\%} of victims migrate to private routing within 60 days, rising to \textbf{54\%} after repeated exposures, yet many appear unaware that private channels themselves can be sandwiched. Churn remains limited: only \textbf{7.5\%} of users fail to reactivate after their first sandwich, so most continue transacting and become persistent targets. At the same time, private routing grows from \textbf{31.8\%} of all transactions in November 2024 to \textbf{50.1\%} in February 2025, suggesting that individual attempts at protection may inadvertently centralize order flow into a smaller set of intermediaries and increase systemic risk.

All estimates in this study should be read as \emph{lower bounds} on private-path sandwiching. Visibility labels depend on MempoolDumpster coverage and may miss some public submissions, while profit and loss figures rely on ZeroMEV valuations rather than exact payoff reconstruction. Our goal is therefore to characterize structural patterns and scales, not to provide perfectly precise totals.

These results indicate that secure execution cannot rely solely on private order flow channels. More durable protection will likely require protocol level mechanisms that limit extractability, such as commit and reveal schemes, or encrypted intents, together with careful evaluation of emerging architectures such as TEE based.

\bibliographystyle{splncs04}
\bibliography{bibliography}

\end{document}